\documentstyle[prd,aps]{revtex}

\begin{document}
\draft

\twocolumn[
\hsize\textwidth\columnwidth\hsize\csname
@twocolumnfalse\endcsname

\title{\noindent {\small USC-97/HEP-B5\hfill \hfill hep-th/9707215}
\newline
{\small  CERN-TH/97-181 \hfill .}\newline
Superstrings with New Supersymmetry \\
in (9,2) and (10,2) Dimensions}
\author{ \bf Itzhak Bars  and Cemsinan Deliduman\medskip }
\address{\it Theory Division, CERN, CH-1211 Geneva 23, Switzerland \\
and \\
Department of Physics and Astronomy, \\
University of Southern California, Los Angeles, CA 90089-0484}
\date{July 25, 1997}
\maketitle
\begin{abstract}
We construct superstring theories that obey the new supersymmetry algebra 
\{${Q_\alpha , Q_\beta}$\}$=\gamma_{\alpha \beta}^{\mu \nu} P_{1\mu}P_{2\nu}$ in
a Green-Schwarz formalism, with kappa supersymmetry also of the new type.
The superstring is in a system with a superparticle so that their total momenta are
$P_{2\mu},P_{1\nu}$ respectively. The system is covariant and critical in (10,2) 
dimensions if the particle is massless and in (9,2) dimensions if the particle is massive.
Both the superstring and superparticle have coordinates with two timelike dimensions
but each behaves effectively as if they have a single timelike dimension. This is due
to gauge symmetries and associated constraints. We show how to generalize the
gauge principle to more intricate systems containing two parts, 1 and 2. 
Each part contains interacting constituents, such as p-branes, and each part behaves
effectively as if they have one timelike coordinate, although the full system has
two timelike coordinates. The examples of two superparticles, and of a superparticle 
and a superstring, discussed in more detail are special cases of such a 
generalized interacting system. 
\end{abstract}
\vskip2pc
]

\section{Introduction}

The duality properties of superstring and super p-brane theories have led to
hints that there may be hidden timelike dimensions in the underlying theory 
\cite{ibtokyo}-\cite{periwal}. Unifying dualities, supersymmetry and Lorentz
invariance suggests a structure in 14 dimensions with signature (11,3)
including three timelike dimensions \cite{d14}-\cite{typeB}. For simplicity
one may concentrate on two timelike dimensions since the conceptual jump
from one to two timelike dimensions can be easily extended to three. At
issue is how to make sense of the extra timelike dimensions?

A main feature of this setting is also that the supersymmetry deviates
substantially from the standard one, while its structure is intimately
related to the answers we seek. By constructing explicit models with the new
supersymmetry, and with two timelike dimensions, progress was made in
relating these concepts to the familiar one-time world with standard
supersymmetry.

In the present paper more progress is reported by constructing superparticle
and superstring models in $(d-2,2)$ dimensions. In the first part of the
paper we will discuss a general setup for discussing the two time issues for
two groups of particles or p-branes, with interactions within each group. In
the second and main part of the paper we will construct actions for the
individual superparticles or superstrings as a member of either group.

In recent papers it has been shown that it is possible to formulate an
action principle for particles \cite{twotimes}, extended objects \cite
{partstring}, and field theories \cite{newsusy}, with two timelike
dimensions, without the traditional problems. The main new ingredient is a
gauge symmetry that permits the removal of one timelike dimension for each
particle (or point in an extended object), while maintaining covariance in
the higher space including the extra timelike dimension.

The timelike dimension that is removed is not the same one for the entire
system. In the simplest case there are two particles, labelled \#1 and \#2,
with timelike (or lightlike) momenta $p_1^\mu ,p_2^\mu $ respectively, where 
$\mu =0^{\prime },0,1,2,\cdots ,(d-2),$ is an index in $\left( d-2,2\right) $
dimensions. The position coordinates are $x_1^\mu ,x_2^\mu .$ Each
coordinate has two timelike dimensions labelled by $\mu =0^{\prime },0$. The
gauge invariance gives the constraint $p_1\cdot p_2=0$, and removes the
timelike dimensions $p_1\cdot x_2$ and $p_2\cdot x_1$ in a way that does not
break the SO$\left( d-2,2\right) $ symmetry of the total system. Then
particle \#1 moves effectively with one timelike dimension, and similarly
particle \#2 moves effectively with another orthogonal timelike dimension.

The definition of canonical variables, and quantization, is straightforward
because the formulation involves a single worldline parameter $\tau $ that
appears in $x_1^\mu \left( \tau \right) ,x_2^\mu \left( \tau \right) $. One
should not be confused about trying to introduce two Hamiltonians. It was
shown in \cite{twotimes} that SO$\left( d-2,2\right) $ covariant
quantization is carried out by simply applying the constraints on states.

It was suggested in \cite{twotimes}, that a cosmological scenario is
possible such that, starting at the Big Bang, the particles of type \#2
remain in compactified dimensions, including one timelike dimension, while
particles of type \#1 move in four dimensions in the observed expanding
universe described by an effective theory with a single timelike dimension.

So far the two-times mechanism has been discussed in simple examples without
interactions. In section 2 we expand the scope of this approach by showing
that it also applies to two systems, called \#1 and \#2, each containing any
number of interacting particles or p-branes, with two timelike dimensions.
The {\it total} momenta of each system are $P_{1\mu },P_{2\mu }$
respectively. We show that these replace the momenta $p_1^\mu ,p_2^\mu $ of
individual free particles or p-branes in the mechanism that removes the
extra timelike dimensions. Furthermore, previous discussions had given the
unintended impression that the approach required a new timelike dimension
for each particle or p-brane in the total system. In the present version, it
becomes evident that we can discuss many particles or p-branes with two
timelike dimensions, without needing to add a new timelike dimension for
each new particle or p-brane.

In sections 3--5 we will discuss superparticle and superstring actions that
are supersymmetric under the ``new superalgebra'' 
\begin{equation}
\left\{ Q_\alpha ,Q_\beta \right\} =\gamma _{\alpha \beta }^{\mu \nu
}P_{1\mu }P_{2\nu },\quad P_1\cdot P_2=0.  \label{susy}
\end{equation}
This is the simplest form for a superalgebra in 12D, but it could also be
realized in lower dimensions\footnote{
The dimensions $d$ in which it can be realized are determined by whether $
\gamma _{\alpha \beta }^{\mu \nu }$ is symmetric in $(\alpha \beta )$. These
are $d=\left( 3,4,5\right) \,\,$mod$\left( 8\right) $ assuming a single Weyl
spinor $Q_\alpha $ in the even $4\,$mod$\left( 8\right) $ dimensions. If the
Dirac spinor is taken in even dimensions (i.e. type IIA) then one can also
have $d=6,10$ mod$\left( 8\right) .$ For SO$\left( d-2,2\right) $ the
spinors above are real in these dimensions. Finally, if there is more than
one Weyl spinor (i.e. $Q_{\alpha a}$ $a=1,2,\cdots $) then $\gamma _{\alpha
\beta }^{\mu \nu }$ can be antisymmetric in $(\alpha \beta )$. This allows $
d=7,8,9$ mod$\left( 8\right) .$ The Weyl spinors are complex in these
dimensions, so their complex conjugates must also be added. We will discuss
only the case of symmetric $\gamma _{\alpha \beta }^{\mu \nu }$ in this
paper. The antisymmetric case can be discussed with similar methods. See
footnote (\ref{345}) for a choice of a basis for the gamma matrices.\label
{dims}}. There is also a new local kappa supersymmetry, in a Green-Schwarz
formulation, that parallels the global supersymmetry of this type. We will
show that there are {\it classical} superstring models in ($d-2,2$)
dimensions with type IIA, type IIB or heterotic supersymmetry, that provide
realizations of the new supersymmetry.

The combined superparticle plus superstring system has SO$\left(
d-2,2\right) $ invariance\footnote{
The superalgebra (\ref{susy}) first emerged in a BPS sector of S-theory \cite
{stheory}, where it was suggested as a basis for a $\left( 10,2\right) $
covariant supergravity, with 128 bosonic and 128 fermionic bi-local fields $
\phi (x_1^\mu ,x_2^\mu )$. It was later realized in 12D super Yang-Mills 
\cite{nishsezgin}, 12D supergravity \cite{nishino}, and 12D $\left(
2,1\right) $ heterotic strings \cite{martinec}, but only for a frozen
constant lightlike momentum $P_2^\mu $ which breaks the $\left( 10,2\right) $
covariance. These realizations can be interpreted to be a Kaluza-Klein
reduction in the coordinate $x_2^\mu $ of a bi-local theory. In contrast, an
operator version of both $P_1^\mu ,P_2^\mu $ in the form of derivatives was
realized in bi-local supersymmetric field theory \cite{newsusy}, but in a
compactified version such that $P_1^\mu ,P_2^\mu $ point in orthogonal
directions. In the present paper both $P_1^\mu ,P_2^\mu $ are quantum
operators, and the full SO $\left( d-2,2\right) $ covariance is maintained.}
and a number of gauge invariances. After fixing a subset of gauges and using
a subset of equations of motion, {\it if the particle is frozen} to a fixed
momentum $p_1^\mu ,$ which is either timelike or lightlike depending on the
dimension, then it can be shown that the string reduces to the standard {\it 
classical} supersymmetric string in 3,4,6,10 dimensions. If the particle is
not frozen, the system is more general and maintains the SO$\left(
d-2,2\right) $ invariance.

In the {\it quantum} theory of the combined superparticle and superstring
system, the critical dimension is (10,2) when the superparticle is massless,
and (9,2) when the superparticle is massive (assuming there are no spectator
degrees of freedom).

\section{Two systems and two times}

In \cite{partstring} the general scheme for constructing the action with two
or more timelike dimensions was given. The discussion involved two or more
particles (or p-branes) such that each additional particle (or p-brane)
introduced an additional timelike dimension. In this section we extend this
construction to any number of interacting particles (or p-branes) in system
\#1 plus any number of interacting particles (or p$^{\prime }$-branes) in
system \#2, but without needing to introduce additional timelike dimensions
beyond two. It is straightforward to add system \#3 with a third timelike
dimension, but here we will limit ourselves to only two timelike dimensions.

The generalized multiparticle prescription for the action with two timelike
coordinates is similar to the previous discussion 
\begin{equation}
S=S_1\left( \lambda _2\right) +S_2\left( \lambda _1\right) +\lambda _1\cdot
\lambda _2.  \label{s12}
\end{equation}
However, now $S_1\left( \lambda _2\right) $ describes many particles or
p-branes in system \#1, moving in a background provided by system \#2, where
the background is represented by the moduli $\lambda _2^\mu $. This is a
constant vector constructed from canonical variables of system \#2, and is
independent of worldsheet or worldline parameters. Similarly, $S_2\left(
\lambda _1\right) $ describes particles or p$^{\prime }$-branes in system
\#2, moving in a background provided by system \#1, represented by the
moduli $\lambda _1^\mu $. The constituents of system \#1, labelled by $
i=1,2,\cdots ,N$, may be different ``beasts'' than the constituents of
system \#2 labelled by $j=1,2,\cdots ,M$, and each system may include any
set of interacting p-branes.

The vectors $\lambda _1^\mu ,\lambda _2^\mu $ are directions in which
timelike coordinates will be removed by a gauge principle that is analogous
to the gauged WZW mechanism discussed in \cite{twotimes,partstring}. This
implies covariant derivatives on the branes $x_{1i}^\mu \left( \tau ,\vec{
\sigma}\right) ,x_{2j}^\mu \left( \tau ,\vec{\sigma}^{\prime }\right) $ 
\begin{eqnarray}
D_\alpha x_{1i}^\mu &=&\partial _\alpha x_{1i}^\mu -\lambda _2^\mu
A_{1\alpha i},  \label{cov} \\
D_\alpha x_{2j}^\mu &=&\partial _\alpha x_{2j}^\mu -\lambda _1^\mu
A_{2\alpha j}.  \nonumber
\end{eqnarray}
where $\partial _\alpha ,A_{1\alpha },A_{2\alpha }$ are one-forms on the
brane worldvolumes. The gauge invariance for system \#1 is 
\begin{equation}
\delta x_{1i}^\mu =\lambda _2^\mu \Lambda _{1i},\quad \delta A_{1\alpha
i}=\partial _\alpha \Lambda _{1i},
\end{equation}
and similarly for system \#2. Note that there is a single $\lambda _2^\mu $
independent of $i$, and a single $\lambda _1^\mu $ independent of $j.$

There may be unspecified interactions within each system independently of
the other, but the only interaction considered so far between system \#1 and
system \#2 is the term $\lambda _1\cdot \lambda _2.$ The moduli $\lambda
_1^\mu ,\lambda _2^\mu $ (which are independent of $\tau ,\vec{\sigma}$) are
integrated in the path integral, like other dynamical variables. This
implies that the classical action is minimized with respect to the moduli,
resulting in the equations of motion for $\lambda _1^\mu ,\lambda _2^\mu $, 
\begin{equation}
\lambda _1^\mu =-\frac{\partial S_1\left( \lambda _2\right) }{\partial
\lambda _{2\mu }},\quad \lambda _2^\mu =-\frac{\partial S_2\left( \lambda
_1\right) }{\partial \lambda _{1\mu }}.  \label{lam}
\end{equation}
It will now be shown that $\lambda _1^\mu $ ($\lambda _2^\mu $) is
proportional to the {\it total} momentum $P_1^\mu $ ($P_2^\mu $) of system
\#1 (\#2), for any interactions among the constituents within each system,
and that 
\begin{equation}
P_1\cdot P_2=0,
\end{equation}
as in the single constituent cases discussed in \cite{twotimes}\cite
{partstring}. To do this use the chain rule\footnote{
Here we assume that the only $\lambda _{1,2}^\mu $ dependence in $
S_{1,2}\left( \lambda _{2,1}\right) $ occurs through the covariant
derivatives in the purely bosonic theory. This is not the case in the
supersymmetric theory, as seen in the next section. However, the result is
still the same as it will be shown there (see discussion following eq. (\ref
{lamd}) ).\label{prop}} 
\begin{equation}
-\frac{\partial S_1\left( \lambda _2\right) }{\partial \lambda _{2\mu }}
=\sum_i\int d\tau d\vec{\sigma}\,\frac{\partial S_1\left( \lambda _2\right) 
}{\partial \left( D_\alpha x_{1i}^\mu \right) }A_{1i\alpha }\left( \tau , 
\vec{\sigma}\right) .  \label{lam2}
\end{equation}
Then use the equation of motion for $A_{1i\alpha }\left( \tau ,\vec{\sigma}
\right) $ 
\begin{equation}
\lambda _2^\mu \frac{\partial S_1\left( \lambda _2\right) }{\partial \left(
D_\alpha x_{1i}^\mu \right) }=0\,\,\,\,\rightarrow \lambda _2\cdot D^\alpha
x_{1i}=0,  \label{a1}
\end{equation}
and combine it with (\ref{lam},\ref{lam2}) to show 
\begin{equation}
\lambda _1\cdot \lambda _2=0.  \label{ortho}
\end{equation}
From (\ref{cov},\ref{a1}) one obtains 
\begin{equation}
\lambda _2^2\,\,A_{1i\alpha }\left( \tau ,\vec{\sigma}\right) =\lambda
_2\cdot \partial _\alpha x_{1i}.
\end{equation}
Using the gauge freedom $\Lambda _{1i}\left( \tau ,\vec{\sigma}\right) $ one
may choose the gauge 
\begin{equation}
\lambda _2\cdot x_{1i}\left( \tau ,\vec{\sigma}\right) =a_1\lambda _2^2\,\tau
\end{equation}
where $a_1$ is a constant independent of the particle $i$. In this gauge $
A_{1i\alpha }\left( \tau ,\vec{\sigma}\right) $ simplifies 
\begin{equation}
A_{1i\alpha }\left( \tau ,\vec{\sigma}\right) =a_1\,\partial _\alpha \tau
=a_1\text{ }\delta _{\alpha \tau }.
\end{equation}
Inserting this form in (\ref{lam2}) shows that only the $\alpha =\tau $ term
survives and $\lambda _1^\mu $ becomes\footnote{
A gauge independent way of proving the same result uses the equations of
motion in any gauge, and combines it with the periodicity of the closed
p-branes in the worldsheet variables $\vec{\sigma}.$} 
\begin{eqnarray}
\lambda _1^\mu &=&a_1\sum_i\int d\tau d\vec{\sigma}\,\,p_{1i}^\mu \left(
\tau ,\vec{\sigma}\right) \,=a_1\int d\tau \,\sum_i\,p_{1i}^\mu \left( \tau
\right)  \nonumber \\
&=&a_1P_1^\mu \int_0^Td\tau =a_1T\,P_1^\mu .
\end{eqnarray}
where the definitions of canonical momenta $p_1^{i\mu }\left( \tau ,\vec{
\sigma}\right) =\partial S_1/\partial \left( D_\tau x_{1i}^\mu \right) ,$
center of mass momentum $p_{1i}^\mu \left( \tau \right) =\int d\vec{\sigma}
\,\,p_{1i}^\mu \left( \tau ,\vec{\sigma}\right) \,$ and total conserved
momentum $P_1^\mu =\sum_i\,p_{1i}^\mu \left( \tau \right) $ have been used.
Note that even if there are interactions within system \#1, the total $
P_1^\mu $ is conserved due to translation invariance, although the center of
mass momenta of the various p-branes $p_{1i}^\mu \left( \tau \right) $ may
be time dependent. Thus, $\lambda _1^\mu $ is proportional to the total $
P_1^\mu ,$ and similarly $\lambda _2^\mu $ is proportional to $P_2^\mu .$
These are orthogonal to each other $P_1\cdot P_2=0$ because of (\ref{ortho}
), as promised above.

There is an analog of the positive energy condition for the combined system.
Note that the sign of 
\begin{equation}
P_1^0P_2^{0^{\prime }}-P_2^0P_1^{0^{\prime }}  \label{positive}
\end{equation}
cannot change under SO$\left( d-2,2\right) $ transformations. Combined with
the condition $P_1\cdot P_2=0,$ this implies that the signs of $
P_1^0,P_2^{0^{\prime }}$ are correlated with each other in definite
representations of supersymmetry and SO$\left( d-2,2\right) $. Therefore, if 
$\lambda _1^0\lambda _2^{0^{\prime }}-\lambda _2^0\lambda _1^{0^{\prime }}$
is taken positive in the original action (\ref{s12}) then the combined
system is considered to be in the ``particle-particle sector'' (as opposed
to particle-antiparticle sector) as explained in \cite{twotimes}.

This result extends the scope of the approach in \cite{partstring} to
interacting systems involving any number of constituents within each system
\#1,\#2. The essential points made in \cite{twotimes} about the meaning of
two timelike dimensions, and the cosmological scenario, remain the same as
in the simple two particle case discussed there.

It is straightforward to extend this discussion to three systems
\#1,\#2,\#3, with three timelike dimensions, as in \cite{partstring}, by
allowing many interacting constituents in each system. More than three
timelike dimensions are \`{a} priori possible, but are not needed in our
program since minimal unification of supersymmetry and duality is possible
in 14D with signature $\left( 11,3\right) \,\,$\cite{d14}. Furthermore, more
than three timelike dimensions seem to be forbidden in some supersymmetric
dynamics \cite{sezgin3}.

\section{Superparticles in (d-2,2) dimensions}

We are interested in constructing models with the new supersymmetry of eq. ( 
\ref{susy}) in various dimensions. Massless superparticles in more than
eleven dimensions have been discussed recently in a formalism involving
multiple worldline parameters $\tau _i$ \cite{sezgin3}. To avoid
inconsistencies in the transformation laws with multiple $\tau _i$, and also
be consistent with our generalized scheme (\ref{s12}), we present the
following formulation involving a single worldline parameter $\tau $.
Although there is considerable overlap with (\ref{s12}) for the single
particle in a background of the other, there are differences in the
supersymmetry transformation rules. Furthermore, we also discuss massless as
well as massive superparticles, superparticles in lower dimensions, and
superparticles with multiple supersymmetries.

\subsection{Massless superparticle in $\left( d-2,2\right) $}

Consider a massless superparticle \#1 in the background of system \#2. The
background consists of $\lambda _2^\mu $ that can be shown to be
proportional to the center of mass momentum of system \#2, as in the
previous section, and below. The $\tau $ reparametrization invariant action
is $S_1=\int d\tau \,L_1\left( \tau \right) $, with 
\begin{equation}
L_1\left( \tau \right) =\frac 12e_1^{-1}\pi _1^\mu \pi _1^\nu \eta _{\mu \nu
},  \label{l1}
\end{equation}
where 
\begin{equation}
\pi _1^\mu =\dot{x}_1^\mu -\lambda _2^\mu A_1+\bar{\theta}_1\gamma ^{\mu \nu
}\dot{\theta}_1\,\lambda _{2\nu }.
\end{equation}
The signature is $\left( d-2,2\right) ,\,\,\,\eta _{\mu \nu }=\left(
-,-,+,\cdots ,+,+\right) ,$ and labelled by the index $\mu =0^{\prime
},0,1,\cdots ,d-2,$ on $x_1^\mu \left( \tau \right) .$ The spinor
representation for SO$\left( d-2,2\right) $ is labelled by the index $\alpha 
$ on the fermion $\theta _1^\alpha \left( \tau \right) .$ We consider the
dimensions $d$ and the spinor representations in which $\gamma _{\alpha
\beta }^{\mu \nu }$ is symmetric in $(\alpha \beta )$, since this is needed
for the new supersymmetry (\ref{susy}). The following discussion is valid
for all dimensions $d$ that satisfy these criteria\footnote{
As in footnote (\ref{dims}), in $d=\left( 3,4,5\right) $ mod$\left( 8\right) 
$ dimensions, with $\mu =0,1,2,\cdots ,0^{\prime },$ the real (Weyl) spinor
has dimension $2^{\left[ \frac{d-1}2\right] }$, and the real gamma matrices
are given by $\gamma ^\mu =(\gamma ^m,\gamma ^{0^{\prime }}),$ where the
last one is proportional to the identity $(\gamma ^{0^{\prime
}})_{\,\,\,\beta }^\alpha =\delta _{\,\,\beta }^\alpha ,$ while the other $
\gamma ^m$ are real and traceless. For $d=6,10$ mod$\left( 8\right) $ the
real Dirac spinor has dimension $2^{d/2}$ and the real gamma matrices in
these dimensions can again be taken as above, with $\gamma ^{0^{\prime }}=1.$
Then $\gamma ^{\mu \nu }=(\gamma ^{mn},\gamma ^{m0^{\prime }})$ are defined
by $\gamma ^{m0^{\prime }}=\gamma ^m$ and $\gamma ^{mn}=\frac 12[\gamma
^m,\gamma ^n].$ These form the algebra of SO$\left( d-2,2\right) .$
Explicitly, for $d=3:\,(\gamma ^\mu )_{\,\,\,\beta }^\alpha =\left( i\sigma
_2,\sigma _3,1\right) $ $;$ for $d=4:\,$ $(\gamma ^\mu )_{\,\,\,\beta
}^\alpha =(i\sigma _2,\sigma _{1,}\sigma _3,1)\,;$ for $d=5:\,$ $(\gamma
^\mu )_\alpha ^{\,\,\,\beta }=(i\sigma _2\otimes \sigma _1,\sigma _1\otimes
\sigma _1,\sigma _3\otimes \sigma _1,1_2\otimes \sigma _3,1_2\otimes 1_2)$.
When multiplied by the antisymmetric charge conjugation matrix $C=\gamma ^0$
to lower the indices, $C\gamma ^m$ and $C\gamma ^{mn}$ are symmetric. These
symmetric $\gamma _{\alpha \beta }^{\mu \nu }$ are the ones that appear in
the superalgebra, as well as in expressions such as $\bar{\varepsilon}\gamma
^{\mu \nu }\theta $, etc.. The structure is similar for the other
dimensions. \label{345}}. The main interest in this paper is in $\left(
10,2\right) $ and $\left( 9,2\right) $ dimensions since these will turn out
to be the critical dimensions for the string (but not for the particle by
itself). Note in particular that for $\left( 10,2\right) $ the Majorana-Weyl
spinor with 32 real components satisfies this property. The same real spinor
is also the Dirac spinor in $\left( 9,2\right) $ dimensions.

The global supersymmetry transformations that are consistent with the new
superalgebra (\ref{susy}) have the form 
\begin{equation}
\delta _{\varepsilon _1}\theta _1=\varepsilon _1,\quad \delta _{\varepsilon
_1}x_1^\mu =-\bar{\varepsilon}\gamma ^{\mu \nu }\theta _1\lambda _{2\nu
},\quad \delta _{\varepsilon _1}A_1=0.  \label{susy2}
\end{equation}
Under these 
\begin{equation}
\delta _{\varepsilon _1}\pi _1^\mu =0,\quad \delta _{\varepsilon
_1}e_1=0,\quad \delta _{\varepsilon _1}L_1=0.
\end{equation}
There is also local kappa supersymmetry given by 
\begin{eqnarray}
\delta _{\kappa _1}\theta _1 &=&\gamma ^{\mu \nu }\kappa _1\,\,\pi _{1\mu
}\lambda _{2\nu },\quad \delta _{\kappa _1}x_1^\mu =\delta _\kappa \bar{
\theta}_1\gamma ^{\mu \nu }\theta _1\,\lambda _{2\nu }  \label{kappa} \\
\delta _{\kappa _1}A_1 &=&2\lambda _2\cdot \pi _1\,\bar{\kappa}_1\dot{\theta}
_1,\quad \delta _{\kappa _1}e_1^{-1}=4\lambda _2^2\,\bar{\kappa}_1\dot{\theta
}_1\,\,e_1^{-1}.  \nonumber
\end{eqnarray}
Under the local $\kappa _1\left( \tau \right) $ supersymmetry the Lagrangian
is invariant.

Note that we have used $\delta _{\varepsilon _1}\lambda _2^\mu =\delta
_{\kappa _1}\lambda _2^\mu =0$ in proving that $S_1\left( \lambda _2\right) $
is supersymmetric. Similarly, in verifying the supersymmetry of system \#2, $
S_2\left( \lambda _1\right) ,$ we need to use $\delta _{\varepsilon
_1}\lambda _1^\mu =\delta _{\kappa _1}\lambda _1^\mu =0.$ Since we expect
that $\lambda _1\sim p_1^\mu $ after using the equations of motion, we need
to check if $\delta _{\varepsilon _1}p_1^\mu =\delta _{\kappa _1}p_1^\mu =0$
on shell, for consistency. This will be verified following eq.(\ref{tdot}).

The momentum is 
\begin{equation}
p_1^\mu =e_1^{-1}\pi _1^\mu .
\end{equation}
In the first order formalism the action may be rewritten as 
\begin{equation}
S_1=\int d\tau \left( 
\begin{array}{c}
p_1\cdot \dot{x}_1-p_1\cdot \lambda _2\,A_1 \\ 
+\bar{\theta}_1\gamma ^{\mu \nu }\dot{\theta}_1\,p_{1\mu }\lambda _{2\nu }-
\frac 12e_1p_1^2
\end{array}
\right) .  \label{first}
\end{equation}
The supersymmetry of this form may be shown by deriving the transformation
laws for the momentum from its definition. We find that $\delta
_{\varepsilon _1}p_1^\mu =0,$ but $\delta _{\kappa _1}p_1^\mu $ is non-zero
off shell (unlike \cite{sezgin3}) 
\begin{equation}
\delta _{\kappa _1}p_1^\mu =2\bar{\kappa}_1\left( 
\begin{array}{c}
\lambda _2^2\gamma ^\mu \,\gamma \cdot p_1\ -\gamma \cdot \lambda
_2\,\,\gamma \cdot p_1\,\lambda _2^\mu  \\ 
+\lambda _2\cdot p_1\,\,\gamma \cdot \lambda _2\,\,\gamma ^\mu \,
\end{array}
\right) \dot{\theta}_1.  \label{dp}
\end{equation}

The constraints and equations of motion that follow from (\ref{first}) or ( 
\ref{l1}) are 
\begin{eqnarray}
p_1^2 &=&0,\quad \lambda _2\cdot p_1=0,\quad \\
\dot{p}_1 &=&0,\quad \left( \gamma ^{\mu \nu }\dot{\theta}_1\right) p_{1\mu
}\lambda _{2\nu }=0.  \nonumber
\end{eqnarray}
and the equation of motion for $\lambda _2$ in (\ref{s12}) gives 
\begin{equation}
\lambda _1^\mu =\int d\tau \left( A_1p_1^\mu +\bar{\theta}_1\gamma ^{\mu \nu
}\dot{\theta}_1p_{1\nu }\right) .  \label{lamd}
\end{equation}
The first term is proportional to $p_1^\mu $, which is consistent with $
\lambda _1^\mu \sim p_1^\mu $ as desired, but the second term is not
obviously so (see footnote ($\ref{prop}$) ). The general solution of the
equation of motion for $\theta _1$ is 
\begin{equation}
\dot{\theta}_1=\left( \gamma ^{\mu \nu }\psi \right) \,\,p_{1\mu }\lambda
_{2\nu },  \label{tdot}
\end{equation}
where $\psi \left( \tau \right) $ is any spinor. Inserting this in (\ref
{lamd}) and using the constraints, it follows that the second term in (\ref
{lamd}) is also proportional to $p_1^\mu ,$ for any $\psi \left( \tau
\right) .$ Since $p_{1\mu }$ is time independent it can be pulled out of the
integral to show once again that $\lambda _1^\mu \sim p_1^\mu ,$ on shell,
as expected.

Now we can verify if $\delta _{\kappa _1}\lambda _1^\mu =0$ is consistent
with $\lambda _1^\mu \sim p_1^\mu ,$ on shell. In the off shell expression $
\delta _{\kappa _1}p_1^\mu $ in (\ref{dp}) we insert the general form (\ref
{tdot}) and use the constraints. The result is $\delta _{\kappa _1}p_1^\mu
=0 $ on shell, consistent with the original assumption $\delta _{\kappa
_1}\lambda _1^\mu =0$.

The superparticles with the new supersymmetry (\ref{susy},\ref{susy2},\ref
{kappa}) are consistent both at the classical and quantum levels, in any
number of dimensions $d$ for which $\gamma _{\alpha \beta }^{\mu \nu }$ are
symmetric.

Furthermore, it is possible to have any number of supersymmetries by
appending an index on $\theta _{1\alpha }^A\left( \tau \right) $ and on $
\varepsilon _1^A,\kappa _{1\alpha }^A\left( \tau \right) ,\,\,A=1,2,\cdots N.
$ There are no restrictions on these generalizations, just as for standard
supersymmetry.

The action (\ref{l1}) or (\ref{first}) has one more bosonic local symmetry
with parameter $\omega (\tau )$ 
\begin{eqnarray}
\delta _\omega \theta &=&\omega \dot{\theta},\quad \delta _\omega x^\mu
=-\omega \bar{\theta}\gamma ^{\mu \nu }\dot{\theta}\lambda _\nu ,  \nonumber
\\
\delta _\omega e &=&0,\quad \delta _\omega A_1=0,
\end{eqnarray}
which gives $\delta _\omega \pi ^\mu =0.$ However this symmetry has no
additional implications for the on shell theory other than those of $\kappa $
and reparametrization invariances.

\subsection{Massive superparticle in $\left( d-2,2\right) $}

In the discussion above the superparticle was massless according to the
constraint $p_1^2=0.$ It is easy to obtain an action for the massive
superparticle by doing Kaluza-Klein reduction. For example if the massless
superparticle is in $\left( 10,2\right) $ we reduce the 11th dimension to
arrive to a massive superparticle in $\left( 9,2\right) $. For this we
distinguish one spacelike component, which will be denoted $\mu =11$ in
every dimension (instead of calling it $\mu =d-1$). Then replace the
corresponding momentum with the constant mass parameter $p_1^{11}\equiv m_1$
, while throwing away the corresponding component in $\lambda _2^{11}=0.$ We
thus have one less spacelike dimensions as compared to the massless particle
case discussed above. The first order action (\ref{first}) reduces to 
\begin{equation}
S_1=\int d\tau \left( 
\begin{array}{c}
p_1\cdot \dot{x}_1-p_1\cdot \lambda _2\,A_1 \\ 
+\bar{\theta}_1\gamma ^{\mu \nu }\dot{\theta}_1\,p_{1\mu }\lambda _{2\nu }- 
\frac 12e_1p_1^2 \\ 
+\bar{\theta}_1\gamma ^{11,\nu }\dot{\theta}_1\,m_1\lambda _{2\nu }-\frac 12
e_1m_1^2
\end{array}
\right) .  \label{massive}
\end{equation}
where $\mu ,\nu =0^{\prime },0,1,\cdots ,d-2$. We have dropped $\dot{x}
_1^{11}m_1$ since it is a total derivative. Thus, the compactified
coordinate $x_1^{11}$ drops out. The supersymmetry transformations follow
from the ones in (\ref{susy2},\ref{kappa}) by dropping the direction $\mu
=11 $ in $x_1^\mu ,\lambda _2^\mu $ and using $\delta m_1=0.$ Note however
that now the $\kappa $ transformation has one more term proportional to the
mass 
\begin{equation}
\delta _{\kappa _1}\theta _1=\gamma ^{\mu \nu }\kappa _1\,\,\pi _{1\mu
}\lambda _{2\nu }+\gamma ^{11,\nu }\kappa _1\lambda _{2\nu }\,\,m_1e_1.
\end{equation}
Thus, from the massless superparticle action in $\left( 10,2\right) $ we
derive the massive superparticle action in $\left( 9,2\right) ,$ and
similarly for other values of $d.$ Using the equations of motion it is seen
that the superparticle is massive.

If two massive superparticles are coupled as in the usual scheme (\ref{s12}
), and then we keep the components $\lambda _{1,2}^{11}\sim m_{1,2}$ for
both particles in doing the reduction (note $\lambda _2^{11}=0$ above), then
the massive Dirac operator for the combined system becomes 
\begin{equation}
\left( \gamma ^\mu p_{1\mu }+\gamma ^{11}m_1\right) \left( \gamma ^\mu
p_{2\mu }+\gamma ^{11}m_2\right) ,
\end{equation}
with the constraint 
\begin{equation}
p_1\cdot p_2+m_1m_2=0,
\end{equation}
while the massive Klein Gordon operator that is obtained by squaring the
Dirac operator is, $(p_1^2+m_1^2)(\,p_2^2+m_2^2).$ 

There are two global supersymmetries, since the total action is invariant
under two separate parameters $\varepsilon _{1,2}$ as in (\ref{susy2}).
After using $\lambda _{1,2}^\mu \sim p_{1,2}^\mu $ one finds the $N=2$
superalgebra by applying (\ref{susy2}) twice, i.e. $\left[ \delta
_{\varepsilon _i},\delta _{\varepsilon _j}\right] x_{1,2}^\mu $, for $i,j=1,2
$ gives 
\begin{equation}
\left\{ Q_{i\alpha },Q_{j\beta }\right\} =\delta _{ij}\,\gamma _{\alpha
\beta }^{\mu \nu }\,\,p_{1\mu }\,p_{2\nu }.
\end{equation}
When particle \#2 is frozen, and rotated to a standard frame (massless or
massive), the superalgebra of particle \#1 reduces to the standard $N=1$
supersymmetry. When particle \#2 is allowed to perform all allowed motions
the combined system has two supersymmetries. Similar comments apply to the
superparticle and superstring system discussed in the following sections.

In constructing field theories that describe the combined system \#1 and \#2
(see e.g. \cite{newsusy}) these structures need to be taken into account.

\section{Superstrings in (9,2) dimensions}

We will discuss the system of one superparticle and one superstring as an
example of the more general scheme (\ref{s12}). Since the superparticle is
discussed in the previous section we will concentrate on the superstring
action. To keep the notation simple we will omit the index \#2 on the string
supercoordinates $x_2^\mu \left( \tau ,\sigma \right) ,\theta _{2\alpha
}\left( \tau ,\sigma \right) ,$ and the index \#1 on $\lambda _1^\mu .$ We
first discuss the string in the background of a {\it massive} superparticle,
thus $\lambda ^\mu \sim p_1^\mu $, and therefore $\lambda ^2$ is timelike.
The following discussion applies to $d=4,5,11$ with signature $\left(
d-2,2\right) $, but we are mainly interested in $\left( 9,2\right) .$ The
spacetime index takes the values $\mu =0,1,\cdots ,d-2,0^{\prime },$. The
dimension of the spinor is 2$^{\left[ \frac{d-1}2\right] }$ and the gamma
matrices $\gamma ^\mu $ live in this space as described in footnote (\ref
{345}). There is an additional gamma matrix $\gamma ^{\left( d-1\right) }$
which will be denoted $\gamma ^{11}$ in every dimension for convenience of
notation, as in the previous section on massive superparticles. Note that $
\gamma ^{11}$ is a scalar under SO$\left( d-2,2\right) .$ Furthermore, while 
$\gamma ^{11}$ is traceless for the cases $d=5,11$, it is equal to 1 for the
case $d=4$ since the spinor is already Weyl projected.

The action we propose is 
\begin{eqnarray}
S_2\left( \lambda \right) &=&\frac 12\int d^2\sigma \,\,\sqrt{-g}g^{ij}\pi
_i^\mu \pi _j^\nu \eta _{\mu \nu }  \nonumber \\
&&-\int d^2\sigma \,\,\varepsilon ^{ij}\,\partial _ix^\mu \,\bar{\theta}
\gamma _{\mu \nu }\gamma ^{11}\partial _j\theta \,\,\lambda ^\nu  \label{2a}
\\
&&+\frac{\sqrt{-\lambda ^2}}2\int d^2\sigma \,\,\varepsilon ^{ij}\,\,\,\bar{
\theta}\gamma ^\mu \gamma ^{11}\partial _i\theta \,\,\bar{\theta}\gamma
_{\mu \nu }\partial _j\theta \,\lambda ^\nu  \nonumber
\end{eqnarray}
where 
\begin{equation}
\pi _i^\mu =\partial _ix^\mu -\lambda _1^\mu A_i+\bar{\theta}\gamma ^{\mu
\nu }\partial _i\theta \,\,\lambda _\nu \,.  \label{pi}
\end{equation}
Evidently there is reparametrization invariance, and the following gauge
invariance 
\begin{equation}
\delta x^\mu =\lambda ^\mu \Lambda _2\left( \tau ,\sigma \right) ,\quad
\delta A_i=\partial _i\Lambda _2\left( \tau ,\sigma \right) .  \label{gaug}
\end{equation}

Under the global supersymmetry transformations with the new superalgebra ( 
\ref{susy}) one has 
\begin{equation}
\delta _\varepsilon \theta =\varepsilon ,\quad \delta _\varepsilon x^\mu =- 
\bar{\varepsilon}\gamma ^{\mu \nu }\theta \,\lambda _\nu ,\quad \delta
_\varepsilon A_i=0,\quad \delta _\varepsilon g_{ij}=0.  \label{super}
\end{equation}
This gives 
\begin{equation}
\delta _\varepsilon \pi _i^\mu =0,
\end{equation}
showing that first term of $S_2$ is supersymmetric. The remainder of $
S_2\left( \lambda \right) $ gives 
\begin{eqnarray}
\delta _\varepsilon S_2 &=&\int d^2\sigma \,\,\varepsilon ^{ij}\,\,\,\left( 
\bar{\varepsilon}\gamma _{\mu \rho }\partial _i\theta \,\,\bar{\theta}\gamma
^{\mu \nu }\gamma ^{11}\partial _j\theta \right) \lambda _\nu \lambda ^\rho
\label{identity} \\
&&+\frac{\sqrt{-\lambda ^2}}2\int d^2\sigma \,\,\varepsilon
^{ij}\,\,\,\left( 
\begin{array}{c}
\bar{\varepsilon}\gamma ^\mu \gamma ^{11}\partial _i\theta \,\,\bar{\theta}
\gamma _{\mu \nu }\partial _j\theta \\ 
-\bar{\varepsilon}\gamma _{\mu \nu }\partial _i\theta \,\,\bar{\theta}\gamma
^\mu \gamma ^{11}\partial _j\theta
\end{array}
\right) \lambda ^\nu .  \nonumber
\end{eqnarray}
The integrand of the first term can be rewritten as 
\begin{eqnarray}
&&\frac 13\varepsilon ^{ij}\left( 
\begin{array}{l}
2\bar{\varepsilon}\gamma _{\mu \rho }\partial _i\theta \,\,\bar{\theta}
\gamma ^{\mu \nu }\gamma ^{11}\partial _j\theta \\ 
-\bar{\varepsilon}\gamma _{\mu \rho }\theta \,\,\partial _i\bar{\theta}
\gamma ^{\mu \nu }\gamma ^{11}\partial _j\theta
\end{array}
\right) \lambda _\nu \lambda ^\rho  \label{deriv} \\
&&+\frac 13\varepsilon ^{ij}\partial _i\left( \bar{\varepsilon}\gamma _{\mu
\rho }\theta \,\,\bar{\theta}\gamma ^{\mu \nu }\gamma ^{11}\partial _j\theta
\right) \lambda _\nu \lambda ^\rho .  \nonumber
\end{eqnarray}
Since the last term is a total derivative it is dropped in $\delta
_\varepsilon S_2$. The second term in $\delta _\varepsilon S_2$ can be
written similarly by dropping total derivatives. Then one gets the integrand
for the second term 
\begin{equation}
\frac{\sqrt{-\lambda ^2}\varepsilon ^{ij}}6\left( 
\begin{array}{l}
2\bar{\varepsilon}\gamma ^\mu \gamma ^{11}\partial _i\theta \,\,\,\bar{
\theta }\gamma _{\mu \nu }\partial _j\theta \\ 
-\bar{\varepsilon}\gamma ^\mu \gamma ^{11}\theta \,\,\,\partial _i\bar{
\theta }\gamma _{\mu \nu }\partial _j\theta \\ 
-2\bar{\varepsilon}\gamma _{\mu \nu }\partial _i\theta \,\,\,\bar{\theta}
\gamma ^\mu \gamma ^{11}\partial _j\theta \\ 
+\bar{\varepsilon}\gamma _{\mu \nu }\theta \,\,\,\partial _i\bar{\theta}
\gamma ^\mu \gamma ^{11}\partial _j\theta
\end{array}
\right) \lambda ^\nu .  \label{deriv2}
\end{equation}
One can use $\left( 2,2\right) ,\left( 3,2\right) $ and $\left( 9,2\right) $
gamma matrix identities to show that the two integrands combined gives zero
for any timelike $\lambda ^\mu $. To prove it we take advantage of the SO$
\left( d-2,2\right) $ symmetry of the combined string and particle system ( 
\ref{s12}), and choose a Lorentz frame for the particle by taking $\lambda
^\mu =(0,\vec{0},\lambda ^{0^{\prime }})$. The little group is SO$(d-2,1)$.
It is useful to define chiral spinors of the little group in the remaining $
\left( 2,1\right) ,$ $\left( 3,1\right) \,\,$or $\left( 9,1\right) $
dimensions by $\theta _{\pm }=\frac 12(1\pm \gamma ^{11})\theta $ (note that
for $\left( 2,1\right) $ case $\theta _{-}=0$ since $\gamma ^{11}=1$ for $
d=4 $ in the Weyl sector, as explained above). In this frame the total
integrand becomes 
\begin{equation}
\frac 13(\lambda ^{0^{\prime }})^2\varepsilon ^{ij}\,\,\sum_{\pm }\,\left( 
\begin{array}{l}
2\bar{\varepsilon}_{\pm }\gamma ^m\partial _i\theta _{\pm }\,\,\,\,\bar{
\theta}_{\pm }\gamma _m\partial _j\theta _{\pm } \\ 
-\bar{\varepsilon}_{\pm }\gamma ^m\theta _{\pm }\,\,\,\,\partial _i\bar{
\theta _{\pm }}\gamma _m\partial _j\theta _{\pm }
\end{array}
\right) ,  \label{identity0}
\end{equation}
where $m$ is an index in the remaining $(d-2,1)$ little group dimensions.
Now we can use the well known gamma matrix identities for symmetrized
indices $(\alpha \beta \gamma \delta ),$ $(\gamma ^m)_{(\alpha \beta
}(\gamma _m)_{\gamma \delta )}=0$, valid in $(2,1)$, $\left( 3,1\right) $
and $\left( 9,1\right) $ dimensions, to show that the integrand (\ref
{identity0}) is a total derivative, as in the usual Green-Schwarz
superstring in these little group dimensions. We have thus shown that (\ref
{identity}) vanishes in any frame only for $d=4,5,11$. So, the action is
supersymmetric for any timelike $\lambda ^\mu $ (with appropriate positive
energy conditions\footnote{
Note that in the special frame we have used $\sqrt{-\lambda ^2}=\left|
\lambda ^{0^{\prime }}\right| =\lambda ^{0^{\prime }}$ by recalling the
positive energy condition (\ref{positive}) and the fact that we wish to
describe a superstring with positive energy $P^0\geq 0.$ More generally in
the total action (\ref{s12}) that includes (\ref{2a}) one may multiply $
\sqrt{-\lambda _1^2}$ with the sign of $\lambda _1^0\lambda _2^{0^{\prime
}}-\lambda _2^0\lambda _1^{0^{\prime }}$ to insure the desired result.}) in
these dimensions.

Next we consider local $\kappa $ supersymmetry with 
\begin{eqnarray}
\delta _\kappa x^\mu &=&\delta _\kappa \bar{\theta}\gamma ^{\mu \nu }\theta
\lambda _\nu ,\quad \delta _\kappa \bar{\theta}=\bar{\kappa}^l\pi _l^\mu
\gamma _{\mu \nu }\lambda ^\nu ,  \nonumber \\
\delta _\kappa A_j &=&2\bar{\kappa}^l(\pi _l\cdot \lambda )\left( \partial
_j\theta -\frac{g_{ji}\epsilon ^{ik}\gamma ^{11}}{\sqrt{-g}}\partial
_k\theta \right) , \\
\delta _\kappa (\sqrt{-g}g^{ij}) &=&2\lambda ^2\left( 
\begin{array}{l}
\bar{\kappa}^j\left( \sqrt{-g}g^{il}-\epsilon ^{il}\gamma ^{11}\right)
\partial _l\theta \\ 
+\bar{\kappa}^i\left( \sqrt{-g}g^{jl}-\epsilon ^{jl}\gamma ^{11}\right)
\partial _l\theta
\end{array}
\right) .  \nonumber
\end{eqnarray}
They are constructed so that the two fermi terms in the $\kappa $ variation
of the action vanish. The four fermi terms in the variation are 
\begin{eqnarray}
&&-\,\,\epsilon ^{ij}\left( 
\begin{array}{l}
2\delta _\kappa \bar{\theta}\gamma _{\mu \nu }\gamma ^{11}\partial _i\theta
\,\,\,\bar{\theta}\gamma ^{\mu \rho }\partial _j\theta \\ 
-\delta _\kappa \bar{\theta}\gamma ^{\mu \rho }\theta \,\,\,\partial _i\bar{
\theta}\gamma _{\mu \nu }\gamma ^{11}\partial _j\theta
\end{array}
\right) \lambda ^\nu \lambda _\rho  \label{identity2} \\
&&+\frac{\sqrt{-\lambda ^2}}2\,\epsilon ^{ij}\left( 
\begin{array}{l}
2\delta _\kappa \bar{\theta}\gamma ^\mu \gamma ^{11}\partial _i\theta \,\, 
\bar{\theta}\gamma _{\mu \nu }\partial _j\theta \\ 
+\delta _\kappa \bar{\theta}\gamma ^\mu \gamma ^{11}\theta \,\,\,\partial _i 
\bar{\theta}\gamma _{\mu \nu }\partial _j\theta \\ 
-2\delta _\kappa \bar{\theta}\gamma _{\mu \nu }\partial _i\theta \,\,\,\bar{
\theta}\gamma ^\mu \gamma ^{11}\partial _j\theta \\ 
-\delta _\kappa \bar{\theta}\gamma _{\mu \nu }\theta \,\,\,\partial _i\bar{
\theta}\gamma ^\mu \gamma ^{11}\partial _j\theta
\end{array}
\right) \lambda ^\nu .  \nonumber
\end{eqnarray}
The structure of these terms are similar to those of (\ref{deriv},\ref
{deriv2}) except for the insertions of $\gamma ^{11}$ being in different
places. By using gamma matrix identities in the special dimensions $\left(
2,2\right) ,\,\left( 3,2\right) ,\,\left( 9,2\right) $ it can be shown that
this is a total derivative for any timelike $\lambda ^\mu $. To see this, we
can use once again the special frame and reduce these four fermi terms to
the same structure as (\ref{identity0}) except for replacing $\delta _\kappa 
\bar{\theta}$ instead of $\bar{\varepsilon}.$ The argument for the vanishing
of $\delta _\kappa S_2\left( \lambda \right) =0$ is then similar to $\delta
_\varepsilon S_2\left( \lambda \right) =0$. So, the action is $\kappa $
supersymmetric for any timelike $\lambda ^\mu $ (with appropriate positive
energy conditions) in dimensions $d=4,5,11$.

Next we wish to analyze the degrees of freedom described by the string
action. After using a subset of equations of motion and fixing a subset of
gauges, it can easily be shown that the action (\ref{2a}) reduces to the
string actions in $\left( 2,1\right) ,\left( 3,1\right) $ and $\left(
9,1\right) $ dimensions with a single timelike coordinate and standard
supersymmetry. To see this, we again choose the special frame for the
particle $\lambda ^\mu =\left( 0,\vec{0},\lambda ^{0^{\prime }}\right) $ so
that (see footnote (\ref{345})) 
\begin{eqnarray}
\gamma _\mu \lambda ^\mu &=&-\lambda ^{0^{\prime }},\quad \gamma _{\mu \nu
}\lambda ^\nu =-\lambda ^{0^{\prime }}\gamma _m \\
\pi _i^{0^{\prime }} &=&\partial _ix^{0^{\prime }}-\lambda ^{0^{\prime
}}A_i,\quad \pi _i^m=\partial _ix^m-\lambda ^{0^{\prime }}\bar{\theta}\gamma
^m\partial _i\theta \,\,  \nonumber
\end{eqnarray}
where $m$ is an index in $d-1$ dimensions with signature $\left(
d-2,1\right) .$ Furthermore, by a gauge choice and solving the corresponding
constraint we can eliminate $x^{0^{\prime }}$ and $A_i$. Then $\pi
_i^{0^{\prime }}=0.$ Defining chiral spinors in $d-1$ dimensions $\theta
_{\pm }=\frac 12(1\pm \gamma ^{11})\theta ,$ one can write the first term in
the action (\ref{2a}) in terms of $\pi _i^m$ as 
\begin{equation}
\pi _i^m=\partial _ix^m-\lambda ^{0^{\prime }}\left( \bar{\theta}_{+}\gamma
^m\partial _i\theta _{+}+\bar{\theta}_{-}\gamma ^m\partial _i\theta
_{-}\right)
\end{equation}
(recall $\theta _{-}=0$ in the case of $d=4).$ Similarly, the second term in
the action (\ref{2a}) becomes 
\begin{equation}
\lambda ^{0^{\prime }}\varepsilon ^{ij}\,\partial _ix^m\,\left( \bar{\theta}
_{+}\gamma _m\partial _i\theta _{+}-\bar{\theta}_{-}\gamma _m\partial
_i\theta _{-}\right) .  \nonumber
\end{equation}
Although not needed, $\lambda ^{0^{\prime }}$ can be absorbed into a
renormalization of the spinors. The remaining string action is recognized as
the Green-Schwarz action in $\left( 2,1\right) ,\left( 3,1\right) ,\left(
9,1\right) $ dimensions (see e.g.\cite{GSW}). This gives the full content of
the degrees of freedom in the special frame. The SO$\left( d-2,2\right) $
covariance of the total system of particle and string is valid with these
degrees of freedom, as can be shown by constructing the generators of SO$
\left( d-2,2\right) $ with methods similar to those given in \cite
{partstring}.

\section{Superstrings in (10,2) dimensions}

As usual we have in mind the total system described by (\ref{s12}). To keep
the notation simple we will again omit the index \#2 on the supercoordinates 
$x_2^\mu \left( \tau ,\sigma \right) ,\theta _2^\alpha \left( \tau ,\sigma
\right) $ and the index \#1 on $\lambda ^\mu $. The following discussion
applies to $d=5,12$ with signature $(d-2,2)$, but we will concentrate mainly
on $(10,2)$. The spinor is real and has dimension 4, 32 for $d=5,12$
respectively.

\subsection{Heterotic String in (d-2,2)}

The action we propose is 
\begin{eqnarray}
S_2^{het}\left( \lambda \right) &=&\frac 12\int d^2\sigma \,\,\sqrt{-g}
g^{ij}\pi _i^\mu \pi _j^\nu \eta _{\mu \nu }  \nonumber \\
&&-\int d^2\sigma \,\,\varepsilon ^{ij}\,\partial _ix^\mu \,\bar{\theta}
\gamma _{\mu \nu }\partial _j\theta \,\,\lambda ^\nu ,  \label{heterotic}
\end{eqnarray}
where $\pi _i^\mu $ is given by (\ref{pi}). As before, there is
reparametrization invariance, gauge invariance as in (\ref{gaug}), and new
supersymmetry under which $\delta _\varepsilon \pi _i^\mu =0,$ as in(\ref
{super}). The first term of $S_2^{het}$ is supersymmetric. The remainder of $
S_2$ gives 
\begin{equation}
\delta _\varepsilon S_2^{het}=\int d^2\sigma \,\,\varepsilon
^{ij}\,\,\,\left( \bar{\varepsilon}\gamma _{\mu \rho }\partial _i\theta \,\, 
\bar{\theta}\gamma ^{\mu \nu }\partial _j\theta \right) \lambda ^\nu \lambda
^\rho .  \label{ident1}
\end{equation}
The integrand can be written as 
\begin{equation}
\frac 13\varepsilon ^{ij}\left( 
\begin{array}{l}
2\bar{\varepsilon}\gamma _{\mu \rho }\partial _i\theta \bar{\theta}\gamma
^{\mu \nu }\partial _j\theta \\ 
-\bar{\varepsilon}\gamma _{\mu \rho }\theta \partial _i\bar{\theta}\gamma
^{\mu \nu }\partial _j\theta
\end{array}
\right) \lambda _\nu \lambda ^\rho ,  \label{ident}
\end{equation}
by dropping a total derivative term. One can use 5D or 12D gamma matrix
identities to show that its itegral is zero for $\lambda ^2=0$. To prove it
we will choose a Lorentz frame in 5D or 12D, $\lambda ^\mu =\left( \lambda
^{0^{\prime }},0,\vec{0},\lambda ^{11}\right) $. Using footnote (\ref{dims})
one finds 
\begin{equation}
\gamma ^{\mu \nu }\,\lambda _\nu =\left( \,\,-\lambda ^{11}\gamma
^{11},\gamma ^m\left( \,\lambda ^{11}\gamma ^{11}-\lambda ^{0^{\prime
}}\right) ,\,-\lambda ^{0^{\prime }}\gamma ^{11}\right) ,
\end{equation}
where $m$ is in 3D or 10D. In this frame (\ref{ident}) becomes 
\begin{eqnarray}
&&+\left( \lambda _{0^{\prime }}^2-\lambda _{11}^2\right) \varepsilon
^{ij}\left( 
\begin{array}{l}
2\bar{\varepsilon}\gamma ^{11}\partial _i\theta \,\,\,\bar{\theta}\gamma
^{11}\partial _j\theta \\ 
-\bar{\varepsilon}\gamma ^{11}\theta \,\,\,\partial _i\bar{\theta}\gamma
^{11}\partial _j\theta
\end{array}
\right) \\
&&+\varepsilon ^{ij}\left( 
\begin{array}{l}
2\bar{\varepsilon}\gamma ^m(\lambda ^{11}\gamma ^{11}-\lambda ^{0^{\prime
}})\partial _i\theta \,\,\,\bar{\theta}\gamma _m(\,\lambda ^{11}\gamma
^{11}-\lambda ^{0^{\prime }})\partial _j\theta \\ 
-\bar{\varepsilon}\gamma ^m(\,\lambda ^{11}\gamma ^{11}-\lambda ^{0^{\prime
}})\theta \,\,\,\partial _i\bar{\theta}\gamma _m\,(\lambda ^{11}\gamma
^{11}-\lambda ^{0^{\prime }})\partial _j\theta
\end{array}
\right) .  \nonumber
\end{eqnarray}
The 4 or 32 components of $\theta $ in $(3,2)$ or $(10,2)$, can be rewritten
in terms of 2 or 16 dimensional (chiral) spinors $\theta _{\pm }=\frac 12
(1\pm \gamma ^{11})\theta $ of 3D or 10D, respectively. Then, $\bar{\theta}
\gamma ^{11}\partial _j\theta =\partial _j\left( \bar{\theta}_{-}\theta
_{+}\right) $ is a total derivative, and the term proportional to $\left(
\lambda _{0^{\prime }}^2-\lambda _{11}^2\right) $ drops out. Furthermore, $
\bar{\theta}\gamma _m\partial _j\theta =$ $\bar{\theta}_{+}\gamma _m\partial
_j\theta _{+}+\bar{\theta}_{-}\gamma _m\partial _j\theta _{-}$, and $\bar{
\theta}\gamma _m\gamma ^{11}\partial _j\theta =$ $\bar{\theta}_{+}\gamma
_m\partial _j\theta _{+}-\bar{\theta}_{-}\gamma _m\partial _j\theta _{-},$
and similarly for terms involving $\varepsilon _{\pm }.$ The terms with all
spinors of same chirality vanish by virtue of 3D or 10D gamma matrix
identities for symmetrized indices $(\alpha \beta \gamma \delta ),$ $(\gamma
^m)_{(\alpha \beta }(\gamma _m)_{\gamma \delta )}=0.$ The remaining cross
terms are 
\begin{equation}
-\left( \,\lambda _{0^{\prime }}^2-\lambda _{11}^2\right) \sum_{\pm
}\varepsilon ^{ij}\left( 
\begin{array}{l}
2\bar{\varepsilon}_{\pm }\gamma ^m\partial _i\theta _{\pm }\,\,\,\bar{\theta}
_{\mp }\gamma _m\partial _j\theta _{\mp } \\ 
-\bar{\varepsilon}_{\pm }\gamma ^m\theta _{\pm }\,\,\,\partial _i\bar{\theta}
_{\mp }\gamma _m\partial _j\theta _{\mp }
\end{array}
\right) .
\end{equation}
The only solution for $\delta _\varepsilon S\left( \lambda \right) =0$, is $
\lambda _{0^{\prime }}^2-\lambda _{11}^2=-\lambda ^2=0.$

The $\kappa $ symmetry transformations have the same structure as (\ref
{kappa}) except for replacing the insertion $\gamma ^{11}$ by the identity.
The two fermi terms in the $\kappa $ variation of the action vanish. The
four fermi terms are 
\begin{equation}
-\int d^2\sigma \epsilon ^{ij}\left( 
\begin{array}{l}
2\delta _\kappa \bar{\theta}\gamma _{\mu \nu }\partial _i\theta \bar{\theta}
\gamma ^{\mu \rho }\partial _j\theta \\ 
-\delta _\kappa \bar{\theta}\gamma ^{\mu \rho }\theta \partial _i\bar{\theta}
\gamma _{\mu \nu }\partial _j\theta
\end{array}
\right) \lambda ^\nu \lambda _\rho .
\end{equation}
This has the same structure as (\ref{ident}) and thus is zero for $\lambda
^2=0$. Therefore the four fermi term vanishes and the action we proposed for
the $(3,2)$ or $(10,2)$ string is $\kappa $ symmetric for $\lambda ^2=0$.

To determine the degrees of freedom of the string we will partially choose
gauges and use a subset of equations of motion to show that the action (\ref
{heterotic}) reduces to the 3D string and heterotic string in 10D for $
d=5,12 $ respectively, provided $\lambda ^2=0.$ For this we choose a Lorentz
frame $\lambda ^\mu =\left( \lambda ^{0^{\prime }},0,\vec{0},\lambda
^{11}\right) $ with $\lambda ^{0^{\prime }}=\left| \lambda ^{11}\right| ,$
and use lightcone components involving the extra timelike coordinate $
\lambda ^{\pm }=(\lambda ^{0^{\prime }}\pm \lambda ^{11})$ so that all
components of $\lambda ^\mu $ except for $\lambda ^{+}$ are zero, $\lambda
^\mu =\lambda ^{+}\delta _{+}^\mu .$ Then 
\begin{eqnarray}
\gamma _\mu \lambda ^\mu &=&\lambda ^{+}\gamma _{+}=\lambda ^{+}\left(
-1+\gamma ^{11}\right) ,\quad \gamma _{\pm }=\mp 1+\gamma ^{11},  \nonumber
\\
\gamma _{-+} &=&-\gamma ^{11},\;\quad \gamma _{m+}=\gamma _m\left( -1+\gamma
^{11}\right) , \\
\pi _i^{+} &=&\partial _ix^{+}-\lambda ^{+}A_i-\lambda ^{+}\bar{\theta}
\gamma ^{11}\partial _i\theta \,,\quad \,\pi _i^{-}=\partial _ix^{-}, 
\nonumber \\
\pi _i^m &=&\partial _ix^m+\bar{\theta}\gamma ^m\left( -1+\gamma
^{11}\right) \partial _i\theta \,\,\lambda ^{+}\,,  \nonumber
\end{eqnarray}
where $m$ is an index in $(2,1)$ or $(9,1)$ dimensions. The second term in $
S_2$ becomes 
\begin{eqnarray}
&&-\varepsilon ^{ij}\,\partial _ix^\mu \,\bar{\theta}\gamma _{\mu \nu
}\partial _j\theta \,\,\lambda ^\nu  \nonumber \\
&=&-\varepsilon ^{ij}\,\partial _ix^m\,\bar{\theta}\gamma _m\left( -1+\gamma
^{11}\right) \partial _j\theta \,\,\lambda ^{+}  \nonumber \\
&&+\varepsilon ^{ij}\,\partial _ix^{-}\,\bar{\theta}\gamma ^{11}\partial
_j\theta \,\,\lambda ^{+}.
\end{eqnarray}
The 4 or 32 components of $\theta $ in $(3,2)$ or $(10,2)$ can be rewritten
in terms of 2 or 16 dimensional (chiral) spinors $\theta _{\pm }=\frac 12
(1\pm \gamma ^{11})\theta $ of 3D or 10D, respectively. Then, $\bar{\theta}
\gamma ^{11}\partial _j\theta =\partial _j\left( \bar{\theta}_{-}\theta
_{+}\right) $ is a total derivative, and the second term above drops out.

Now we choose gauges and use equations of motion as follows: let $x^{+}=0$
by a gauge choice, $\pi _i^{-}=0$ by equation of motion $\delta A_i$, and $
A_i$ is determined by the equation of motion of $\delta x^{-}.$ The
remaining action is the 3D superstring or 10D heterotic superstring
respectively. We have shown that they are embedded covariantly in $(3,2)$ or 
$\left( 10,2\right) $ dimensions respectively provided the particle to which
it is coupled is massless, since $\lambda ^2=0$.

\subsection{Type IIA and IIB superstrings in (10,2)}

Type IIA,B superstrings are constructed by doubling the fermions $\theta
^{A\alpha },$ $A=1,2.$ The discussion applies in $d=5,12$ so that each
spinor is real and has dimension 4,32 respectively. For type-IIB the 12D
chirality of the two spinors are the same while for type IIA they are
opposite. The covariant momentum is 
\begin{equation}
\pi _i^\mu =\partial _ix^\mu -\lambda ^\mu A_i+\bar{\theta}^A\gamma ^{\mu
\nu }\partial _i\theta ^A\,\,\lambda _\nu \,,
\end{equation}
and the action is 
\begin{eqnarray}
&&S^{II}\left( \lambda \right) =\frac 12\int d^2\sigma \,\,\sqrt{-g}
g^{ij}\pi _i^\mu \pi _j^\nu \eta _{\mu \nu }  \nonumber \\
&&-\int d^2\sigma \,\,\varepsilon ^{ij}\,\partial _ix^\mu \,\left( \bar{
\theta}^1\gamma _{\mu \nu }\partial _j\theta ^1-\bar{\theta}^2\gamma _{\mu
\nu }\partial _j\theta ^2\right) \,\,\lambda ^\nu  \label{2b} \\
&&+\int d^2\sigma \,\,\varepsilon ^{ij}\,\,\,\bar{\theta}^1\gamma ^{\mu \rho
}\partial _i\theta ^1\,\,\bar{\theta}^2\gamma _{\mu \nu }\partial _j\theta
^2\,\,\lambda _\rho \lambda ^\nu .  \nonumber
\end{eqnarray}

The supersymmetry transformation has parameters $\varepsilon ^{A\alpha },$
and is given by (\ref{super}), except that the fermions should be doubled.
As usual $\delta _\varepsilon \pi _i^\mu =0,$ and the first term of the
action is invariant. The second term gives 
\begin{equation}
\delta _\varepsilon S_2^{II}=\int d^2\sigma \,\varepsilon ^{ij}\,\left( 
\begin{array}{c}
\bar{\varepsilon}^1\gamma ^{\mu \rho }\partial _i\theta ^1\bar{\theta}
^1\gamma _{\mu \nu }\partial _j\theta ^1 \\ 
+\bar{\varepsilon}^2\gamma ^{\mu \rho }\partial _i\theta ^2\,\bar{\theta}
^2\gamma _{\mu \nu }\partial _j\theta ^2
\end{array}
\right) \lambda _\rho \,\lambda ^\nu .
\end{equation}
This vanishes since it has the same structure as in the heterotic case (\ref
{ident1}), provided $\lambda ^2=0$.

The local $\kappa $ supersymmetry transformations are given by 
\begin{eqnarray}
\delta _\kappa x^\mu &=&\delta _\kappa \bar{\theta}^A\gamma ^{\mu \nu
}\theta ^A\lambda _\nu ,\quad \delta _\kappa \bar{\theta}^A=\bar{\kappa}
^{Al}\pi _l^\mu \gamma _{\mu \nu }\lambda ^\nu \\
\delta _\kappa (\sqrt{-g}g^{ij}) &=&4\lambda ^2\sqrt{-g}\left( 
\begin{array}{l}
P_{-}^{il}\bar{\kappa}^{1j}\partial _l\theta ^1+P_{+}^{il}\bar{\kappa}
^{2j}\partial _l\theta ^2 \\ 
P_{-}^{jl}\bar{\kappa}^{1i}\partial _l\theta ^1+P_{+}^{jl}\bar{\kappa}
^{2i}\partial _l\theta ^2
\end{array}
\right)  \nonumber \\
\delta _\kappa A_j &=&2(\pi _l\cdot \lambda )\left[ 
\begin{array}{l}
\bar{\kappa}^{1l}\left( \partial _j\theta ^1-\frac 1{\sqrt{-g}}
g_{ji}\epsilon ^{ik}\partial _k\theta ^1\right) \\ 
+\bar{\kappa}^{2l}\left( \partial _j\theta ^2-\frac 1{\sqrt{-g}}
g_{ji}\epsilon ^{ik}\partial _k\theta ^2\right)
\end{array}
\right] .  \nonumber
\end{eqnarray}
where we defined the projection tensors as 
\begin{equation}
P_{\pm }^{ij}=\frac 12\left( g^{ij}\pm \frac{\epsilon ^{ij}}{\sqrt{-g}}
\right) .
\end{equation}
The $\kappa ^A$ parameters are anti-self-dual for A=1 and self-dual for A=2 
\cite{GSW} 
\begin{equation}
\kappa ^{1i}=P_{-}^{ij}\kappa _j^1,\quad \kappa ^{2i}=P_{+}^{ij}\kappa _j^2.
\end{equation}

The $\kappa $ transformations are designed so that two fermi terms in the
variation of the action vanish. The four fermi terms take the form 
\begin{equation}
\epsilon ^{ij}\left( 
\begin{array}{l}
-2\delta _\kappa \bar{\theta ^1}\gamma _{\mu \nu }\partial _i\theta ^1\bar{
\theta ^1}\gamma ^{\mu \rho }\partial _j\theta ^1 \\ 
+\delta _\kappa \bar{\theta ^1}\gamma ^{\mu \rho }\theta ^1\partial _i\bar{
\theta ^1}\gamma _{\mu \nu }\partial _j\theta ^1 \\ 
+2\delta _\kappa \bar{\theta ^2}\gamma _{\mu \nu }\partial _i\theta ^2\bar{
\theta ^2}\gamma ^{\mu \rho }\partial _j\theta ^2 \\ 
-\delta _\kappa \bar{\theta ^2}\gamma ^{\mu \rho }\theta ^2\partial _i\bar{
\theta ^2}\gamma _{\mu \nu }\partial _j\theta ^2
\end{array}
\right) \lambda ^\nu \lambda _\rho .
\end{equation}
This has the same structure as (\ref{ident}) and thus vanishes provided $
\lambda ^2=0$.

To analyze the degrees of freedom we choose gauges and use equations of
motion as in the heterotic case. For $\,\lambda ^2=0$ in $\left( 10,2\right)
,$ this procedure gives the10D superstrings, but now with the fermions
doubled, so that $\theta _{\pm }^{1,2}=\frac 12(1\pm \gamma ^{11})\theta
^{1,2}$ are16 dimensional chiral spinors in $\left( 9,1\right) $. When the
12D chiralities of $\theta ^{1,2}$ are the same/opposite then the 10D
chiralities are also the same/opposite. Therefore we have obtained the IIA,B
superstring theory embedded covariantly in $\left( 10,2\right) $ provided
the particle to which it is coupled is massless, since $\lambda ^2=0.$

The discussion above was at the classical level. The quantum theory can be
analyzed either in SO$\left( d-2,2\right) $ covariant quantization or
lightcone quantization. The methods are discussed in detail in \cite
{partstring}. However, we can quickly determine the outcome since we have
already identified the degrees of freedom and we know the critical
dimensions of the sub-theories. On this basis, we can determine that the
type IIA,B theories (massless particle plus string) are critical and quantum
consistent in (10,2) dimensions. Similarly, the massive particle plus string
theory is critical and quantum consistent in (9,2) dimensions.

The remaining theories in lower dimensions and the heterotic case become
quantum consistent when additional degrees of freedom are added so that the
Virasoro algebra becomes critical.

\end{document}